\documentclass{article}

\usepackage{arxiv}
\usepackage{graphicx}
\usepackage[utf8]{inputenc} 
\usepackage[T1]{fontenc}    
\usepackage{hyperref}       
\usepackage{url}            
\usepackage{booktabs}       
\usepackage{amsfonts}       
\usepackage{nicefrac}       
\usepackage{microtype}      
\usepackage{lipsum}

\title{Is crime concentrated or are we simply using the wrong metrics?}

\author{ Rafael Prieto Curiel \thanks{website: \href{https://rafaelprietocuriel.wordpress.com/}{https://rafaelprietocuriel.wordpress.com/} and \href{https://twitter.com/rafaelprietoc}{Twitter @rafaelprietoc} }\\
Mathematical Institute, University of Oxford\\
Andrew Wiles Building, Radcliffe Observatory Quarter\\
OX2 6GG, Oxford, UK\\
\texttt{rafael.prietocuriel@maths.ox.ac.uk}
}

\begin{document}
\maketitle

\begin{abstract}
Crime is highly concentrated in a few places, is committed by a few offenders and is suffered by a few victims. In recent decades, the concentration of crime has become an accepted fact, yet, little is known in terms of how to measure this concentration of crime such that the metric takes into account the fact that crime has, in general, a low frequency, it fluctuates, it is highly concentrated and has a certain degree of randomness.

Here, the most frequently used metrics for concentration of crimes are reviewed. A null model with complete randomness is used for comparing between different concentration metrics, which allows constructing a sensitivity analysis for every metric against varying crime rates.

Results show that most ways of measuring the concentration of crime are in fact, showing only that crime is rare or that it has fluctuations, but fail to work as a method to compare the concentration of crime between different regions, types of crime or across time.

\end{abstract}


\section{Introduction}

Crime is concentrated in a few places, it is suffered by a reduced number of victims and it is committed by a small number of perpetrators and this seems to be the general pattern observed in crime, regardless of the location being analysed, the time span of the study or the type of crime considered. A similar concentration phenomenon is observed across a wide range of human-related activities (for example, the concentration of wealth or the frequency of family names) and non-human-activities (for example, volcanoes and its eruptions, the location of earthquakes) \cite{NewmanDistributions}.

Things tend to concentrate, and this is commonly known as the ``80/20 rule'', the law of the vital few or the Pareto principle, although the degree in which things concentrate is not the same and plays a relevant role across different disciplines. For instance, the observed income inequality among different countries shows a wide range, from countries with a distribution of income near equity (Kazakhstan or Slovakia) to countries with the highest levels of inequality (South Africa, for example, where the richest 10\% of the population has nearly 32 times the income of the poorest 10\%).

In the case of crime, there are several aspects in which its concentration is analysed: from the time of the day in which crime is committed, the type of crime or the weapons used to perpetrate it, for instance, but the most relevant attributes are 
\begin{itemize}
\item the targets or victims who suffer the crimes \cite{SystematicReviewVictims} and so the counts are commonly the number of crimes suffered by the people, businesses or houses and the concentration is frequently expressed as the percentage of crimes suffered by the most victimised group of observations; 
\item the offenders who commit the crimes \cite{SystematicReviewOffenders} and so the counts are the number of crimes executed by a percentage of the most active criminals, families or criminal groups; and 
\item the places in which crime is executed \cite{SystematicReviewPlaces} where concentration is often expressed in terms of the percentage of street segments or districts in a city with the highest number of crimes.
\end{itemize}

The fact that crime concentrates is described by the \textit{Law of Crime Concentration} \cite{LawCrimeConcentration}, but the degree in which crime is concentrated is a relevant part of the study of crime for several reasons. First, a higher degree of concentration has impacts on policing and crime prevention, since it means that the allocation of resources can be done efficiently, targeting with a higher emphasis the victims, places or potential criminals who are more likely to suffer (or commit) a crime. Secondly, the study of the concentration of crime allows comparing crime between different cities, across different periods, to analyse the concentration between different types of crime; it also allows comparing crime against other social events and determine whether the degree in which crime is concentrated is at all related to the degree in which wealth is concentrated, for instance, and a metric for the concentration of crime helps determine whether an intervention or a security program has an impact displacing crime to other regions or victims \cite{ConcentrationOfCrimeRECC}. Finally, the concentration of crime is also a key element on fear of crime \cite{ModellingFearPrieto}, where in general, a lower concentration of crime means that society will experience more fear \cite{PrietoPalgrave2018}.

Although there might be apparent similarities between crime and other social activities in terms of its distribution and degree of concentration, there are relevant differences between them, which should also mean a crucial difference in the way their concentration should be measured. For instance, in the case of wealth, every person has a certain amount of income and so it is possible to compare the individual income against the average income or against the case in which every person has the same income. Similar to the concentration of wealth or the number of calories consumed, any observed concentration is, mathematically speaking, an observable and stable pattern. It is possible to measure and compare the concentration of wealth through time using standard data techniques, using the Gini coefficient, for instance \cite{GiniFormula, ConcentrationCrimeGini}.

Crime, however, has at least three attributes which make it different to other quantities: crime suffered by individuals is a \emph{variable}, which represent counts (a discrete variable), with a low frequency (almost all observations are equal to zero) and tends to be highly concentrated (and so some observations might be more than one crime). This matters too much since it was found, for instance, that more than 90\% of the British households reported suffering no crime during one year \cite{TseloniBurglary}; less than 10\% of the population is considered to be a ``chronic offender'' \cite{ConcentrationWolfgangOneCohort}; only a small percentage of the people will ever be convicted by the police \cite{YoungDelinquentFerguson}; and a few segments account for the majority of crimes \cite{CrimeConcentrationVaryingCitySize}. Thus, crime is highly concentrated regardless of the dimension in which it is observed. This means that for measuring the concentration of crime, we need to take into account that it is a discrete variable, which might be almost always equal to zero but some observations might have more than one unit (more than one suffered or committed crime).

And why is it relevant to \emph{only} use a correct metric for the concentration of crime? Imagine a metric, $\mu$, any metric, and we apply it naively to the number of crimes in a city, and we observe that our metric $\mu$ is increasing through the years (see, for instance \cite{Pease2018}). Then, we ``accept'' that crime is more concentrated, and so we act against this higher concentration: we design a policy based on this higher concentration; we concentrate too our efforts and resources; we analyse and try to prevent more crimes on the fewer victims. We react against the concentration of crime. But, what if this is wrong? what if we are only using the wrong metric $\mu$ and crime is just as concentrated as before (or even less!)? what if our policy and efforts are based on the wrong metrics? It would be like research on climate change, but not being able to measure temperature or C02 emissions correctly; or like research on obesity but without measuring a person's weight, level of physical activities and food consumption. Crime science too needs to use its correct metrics.

Here, a review of commonly used metrics is analysed, with specific reasons as to why they fail to work as concentration metrics.

\subsection{Concentration of crime metrics}

Consider $N$ individuals and let $c_i$ by the number of crimes that the $i$-th individual suffered during one year. The population suffered $C = \sum c_i$ crimes. Let $v_i = 1$ if the $i$-th person suffered at least one crime during the same year, and zero otherwise, so that $v_i$ indicates the victims, regardless of whether they suffered one or several crimes. There were $V = \sum v_i$ victims, with $C \leq V$. In general, concentration metrics can be viewed as a function $\mu(c_1, c_2, \dots, c_N) = \mu(\mathbf{c})$ which takes all the observed number of individual crimes ($\mathbf{c} = c_1, c_2, \dots, c_N$) and gives a number in return. 

This number, $\mu(\mathbf{c})$ by itself, is almost useless and needs to be compared against other metric(s) $\mu'(\mathbf{c})$ observed for other regions, types of crime or periods of time, in order to detect any pattern of the concentration of crime \cite{HansRosling}. Simply saying that $\mu(\mathbf{c}) = 0.8$, say, provides little and perhaps misleading information about $\mathbf{c}$. For instance, if a person has a blood glucose level of  $20 mg/dL$, it could be a serious case of hypoglycemia, but only detected when compared to the regular $80 mg/dL$ observed on a healthy person.

\subsubsection{Prevalence and frequency}

Often, the level of concentration either at the victim level \cite{SystematicReviewVictims}, the offender \cite{SystematicReviewOffenders} or the location \cite{SystematicReviewPlaces} is reported in terms of the ``prevalence'' $V$, by the ``frequency'' $C$ and by their rate $H = C/V$, which is the average number of crimes suffered by the victims.

\subsubsection{The top v percent suffers T percent of the crimes}

The most victimised units are considered, with a summary such as ``the top $v\%$ most victimised individuals suffered $\tau\%$ of the crimes''. Although no common agreement exists on the levels of $v\%$, frequently the top 1, 5 or 10\% victimised are reported.

\subsubsection{Entropy}

Entropy and other vector metrics could also be used for measuring the concentration of crime. Let $n_i$ be the number of individuals who suffered $i$ crimes. The entropy $E$ is defined as
\begin{equation}
E(\mathbf{c}) = - \sum_i i \log \frac{n_i}{n}.
\end{equation}

\subsubsection{Lorenz curve and Gini coefficient}

Sorting individuals from the one which suffers the lowest number of crimes, $c_{(1)}$ to the one who suffers the highest amount of crimes $c_{(N)}$ allows constructing the cumulative distribution of the number of crimes. The cumulative distribution, with respect to the total number of individuals $N$, and with respect to the total number of crimes $C$, is called the \emph{Lorenz curve}. That is, the horizontal axis goes from 0 to 100\% of the individuals and the vertical axis goes from 0 to 100\% of the total crimes suffered. The Gini coefficient, $G$ is twice the area between the Lorenz curve and the identity line \cite{GiniFormula}. Values of $G$ closer to zero means a more uniform distribution, whereas values of $G$ closer to one means a higher concentration. The Gini coefficient is frequently used to measure income inequality and has been used to measure crime concentration too \cite{TseloniInequality}.

\subsubsection{Filtering observations and the modified Gini coefficient}

A common technique for measuring the concentration of crime is to filter specific observations, for instance, filtering and keeping only individuals which suffered at least one crime and then to apply the Gini coefficient only to the filtered observations.

Also, the Gini coefficient has been adapted to the case of crime, by considering, instead of the total population $N$, the largest number of victims which could be observed with $C$ crimes, that is, $N' = \min \{N, C \}$. The modified Gini coefficient also goes from values of zero, which means a uniform distribution \emph{among the number of individuals who potentially suffered a crime} and one, which means highly concentrated \cite{ConcentrationCrimeGini}.

\subsubsection{Probabilistic models}

A probabilistic approach to the number of crimes suffered by individuals has been considered as a method for constructing concentration metrics. Either by considering that individuals suffer crime following a Poisson distribution \cite{RECCSciReports} or a negative binomial distribution \cite{MohlerConcentration, PoissonRandomRepeats}, the \emph{rate} at which individuals suffer crime is modelled. Then, a global distribution of the rates is assumed, based on the fact that individuals experience different rates (or ``speeds'') and then, the distribution of the rates is analysed for constructing a concentration metric.

In the case of the \emph{rare event concentration coefficient} $RECC$, the assumption is that different rates are observed from a mixture model of different Poisson rates, which is considered as the \emph{victimisation profile}. The $RECC$ is a metric designed particularly for events with a low frequency \cite{RECCSciReports}, which could be highly concentrated, such as volcanic eruptions (which frequently happen on just a few of the active volcanoes around the world), road accidents \cite{RECCRoadAccidents} and also, crime \cite{ConcentrationOfCrimeRECC}. The $RECC$ is obtained by computing as the Gini coefficient from the distribution of the rates.

The \emph{Poisson-gamma concentration} $PGC$ is obtained by assuming that the individual rates are one observation from a gamma distribution with some shape parameter $k$ and a scale parameter $\theta$. The concentration coefficient $PGC$ is obtained by computing the Lorenz curve and the corresponding Gini coefficient from the gamma distribution of the rates, which depends only on the dispersion parameter $k$ \cite{MohlerConcentration}.

In both cases, the rare event concentration coefficient $RECC$ and the Poisson-gamma concentration $PGC$, the relevant aspect is that it does not take into account the number of crimes suffered by individuals directly, but it assumes that the number of crimes is a realisation from an underlying distribution. Then, it assumes that different rates are detected as observations from a global distribution and finally, it constructs a concentration metric from that distribution. See \cite{MohlerConcentration} for an application in the case of Chicago homicide counts.

\section{Too many metrics: which one should we use?}

Having a handful of different metrics for measuring the concentration of crime, poses one serious challenge to the field: which metric should we use and why? Certain attributes are needed (or at least desirable) for the metric we use. What should the ideal metric $\mu(\mathbf{c})$ reveal about crime?

\begin{itemize}
\item \textbf{Stability} - measuring the concentration of crime under similar conditions, should also reveal a similar metric. If the metric varies drastically with small perturbations of $\mathbf{c}$ then it becomes difficult to asses whether we are observing a peak or a valley of the concentration of crime.
\item \textbf{Comparability in time and in space} - the metric should be comparable through different periods of time and should be comparable also between different regions, even if they have a different population size.
\item \textbf{Independence of the population size and the number of crimes} - the metric should not be dependent on the population size or the number of crimes. Thus, a decrease (or increase) in the number of crimes should not also imply an increase (or decrease) in the metric.
\end{itemize}

\subsection{Testing stability, comparability and independence of crime rates}

It is impossible to test metrics for stability, comparability and independence on real-life scenarios, as the underlying concentration pattern is unknown. Therefore, although one metric might seem to be unstable against small perturbations, it could also reveal a change in the crime patterns.

Instead, consider a simulated population of $N$ individuals, all of which suffer crime following a Poisson distribution with a constant rate $\lambda = c$ and apply the metrics of concentration of crime. Although the individual distribution of crime is not precisely a Poisson distribution and although observations might be correlated, that distribution is frequently used for modelling and simulating global patterns of crime \cite{ModellingFearPrieto}. More importantly, it allows varying crime rates $c$ and population size $N$ by applying the most simple crime model.

Notice that the way the simulated population is constructed, all individuals have exactly the same crime rate $c$ and experience the same potential number of crimes. All individuals are equal. Therefore, one simulation of the population gives us what could be observed in a population where crime is not concentrated at all and where individuals suffer a certain crime rate $c$. This, pattern in which there is no concentration should also be revealed by the metrics we use.

For example, with $c= 1$ and with $N = 1,000,000$ we are considering a city with one million inhabitants in which all individuals suffer crime following a Poisson distribution with rate $\lambda = 1$, so that all individuals expect to suffer one crime each year. What is relevant about simulating crime in such a way is that it takes into account the natural randomness that suffering a crime has. All individuals expect to suffer exactly one crime each year, but many of them (37\%) will not suffer crime, simply due to natural fluctuations and the randomness of crime. Simulating allows considering many populations and detecting what would be observed under different scenarios. For instance, with $c = 2$, still, 14\% of the population suffers no crime during a year, although every person expects to suffer two crimes.

The metrics for the concentration of crime were computed for 10,000 simulated populations, all of which have no concentration of crime (all individuals have the same rate) but a varying crime rate $c$, with values of $c$ very close to zero and values below $c = 1$. Reported is the observed value of every metric and so no smoothing was done and outliers are not removed. The code for simulating populations is available \href{https://github.com/rafaelprietocuriel/HomogeneousCrime.git}{here}.

\section{Results}

The construction of the simulations is such that no concentration of crime should be observed. All individuals on the simulated scenario have exactly the same (low) crime rate and so, observing (or reporting) any concentration indicates that any metric is not perfect. The construction of the ``ideal metric'' reflects this, where varying the crime rates has no impact on the metric, and also, the metric shows no concentration.

\begin{figure}
  \centering
    \includegraphics[width=.98\textwidth]{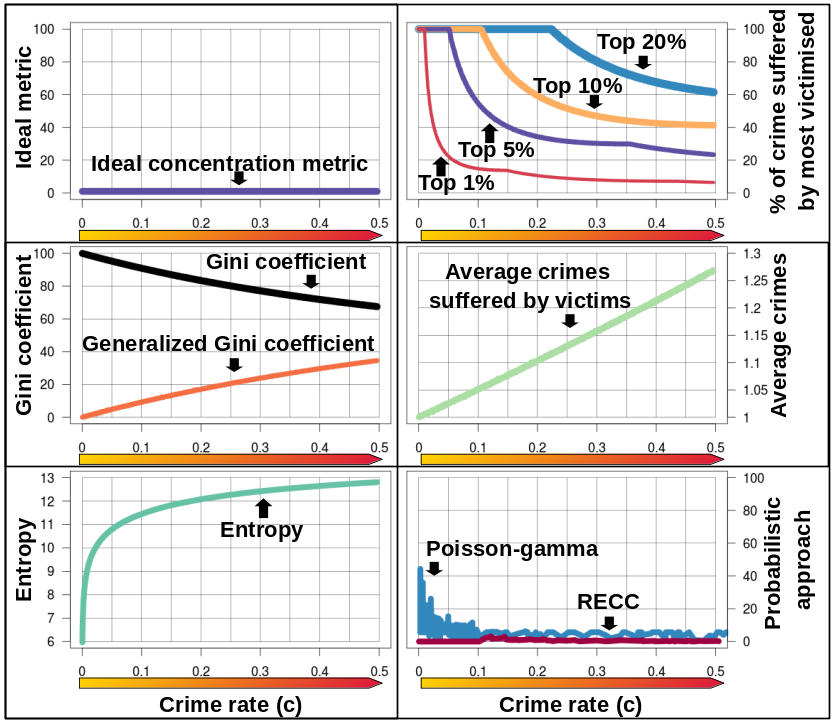}
    \caption{Results of applying each concentration metric to a simulated population in which every individual has exactly the same probability of suffering crimes, thus, the construction itself of the population shows no concentration.}
    \label{ConcentrationMetrics}
\end{figure}

Results of the simulations show that the most frequently used metrics for measuring the concentration of crime have a systematic issue. For example, using the number of crimes suffered by the top 10\% victimised individuals shows that a very small decrease in the crime rates also changes the concentration metric very rapidly (Figure \ref{ConcentrationMetrics}). Also, for a small amount of crime, that concentration metric shows that the top 10\% victimised people suffers all the crime (and so $\mu(\mathbf{c}) = 1$).

The Gini coefficient and the modified Gini coefficient are also affected when crime rates drop and unfortunately, they are affected in the opposite direction. Whilst the Gini coefficient increases with less crime, the modified Gini coefficient decreases with less crime. This actually highlights that in the case of crime concentration, the most frequently used metrics are not even consistent among themselves.

Other metrics, such as the average number of crimes suffered by the victims, or the entropy are also affected by the crime rates.

From the two probabilistic metrics, the Poisson-gamma concentration $PGC$ shows, with higher crime rates, little concentration of crime and a stable metric. However, for lower crime rates, the $PGC$ is unstable and shows an apparent degree of concentration.

The rare event concentration coefficient $RECC$ is, from the explored metrics, the one which is closer to the ideal metric. Although it is not a smooth function, as the other metrics, and does have some uncertainty and roughness, the $RECC$ is not affected by a decrease or increase in the crime rates.

\subsection{Metric stability, comparability and independence of crime rates}

\subsubsection{Prevalence and frequency}

Prevalence and frequency are stable to small perturbations on the crime rates. Both prevalence and frequency are comparable between different types of crime, periods of time or regions. However, there is a seemingly linear decrease of the metric $\mu( \mathbf{c})$ as crime increases.

\subsubsection{The top v percent suffers T percent of the crimes}

The metric, which is perhaps the most frequently used, is not stable to small perturbations or changes in the crime rates. In general, a very small decrease on the crime rates will show a much higher concentration of crime. In essence, when this metric is being used, only a reflection of the crimes rates are obtained.

This metric, however, also has some severe issues, including the lack of agreement on the percentage that gets reported \cite{LorenzCrimeExecuting}; the metric might not be comparable between different cities \cite{CrimeConcentrationVaryingCitySize}; it might be the result of a certain degree of randomness \cite{CrimeConcentrationRecommendations} and it does not work as an adequate metric when the data is extremely sparse. Consider, for example, the number of street segments of The Hague and the number of sexual offences registered by the police between 2007 and 2009 in that city \cite{ConcentrationCrimeGini}. The extremely low frequency of this type of crime (only 430 cases) distributed over the large number of street segments (14,375 segments) means that taking the top 5\% of streets is not even properly defined since, at most, 3\% of the segments concentrate all the crimes. The same issue happens when the top 5\% or top 1\% or any other percentage is considered. Thus, the metric is not comparable between different types of crime, periods of time or regions and is not independent of crime rates.

When we use the top v\% of the victimised individuals and report how much crime they suffer, we might only be measuring a function of crime rates due to the drastic and rapid change that the metric has on the low crime rate regions (Figure \ref{ConcentrationMetrics}).

\subsubsection{Entropy}

Entropy is not stable to small perturbations on the crime rate and it is not independent to the rate. As crime rates increase, the entropy also increases and so it shows an apparent higher concentration of crime. More importantly, with low crime rates, a small increase in the rates show a drastic change in the concentration.

\subsubsection{Lorenz curve and Gini coefficient}

The Gini coefficient is a smooth and stable metric for the concentration of crime, as small increases or decreases do not change drastically its values. However, it is not independent of crime rates. As crime rates increase, the Gini coefficient shows a lower concentration of crime.

\subsubsection{Filtering observations and the modified Gini coefficient}

Although filtering observations seems like a solution to overcome the issues of the high concentration of crime, it has some issues. This technique of considering only crime-involved units, whether they are places or people, and ignoring the units which are not related to crime is a weak solution that, in general, should not be done. In a similar context, it would be like claiming that South Africa has reached income equality when we consider only the richest 10\% of the country. The population or regions that have no crime also provide relevant information in terms of crime concentration that should not be ignored.

Thus, conclusions drawn from this technique, such as ``crime is equally concentrated at different spatial units'' \cite{SystematicReviewComparedToWhat} are fundamentally wrong, particularly when the units which were excluded are 80\% of the addresses, 50\% of the street segments and 20\% of the square grids, or even more, when the same technique is used to ``ignore'' 30\% of the places and more than 50\% of the potential victims and nearly 60\% of the potential offenders to conclude that the concentration of crime among offenders, places and victims is similar \cite{SystematicReviewComparedToWhat}.

\subsubsection{Probabilistic models}

The $PGC$ is close to the ideal metric for higher crime rates but is unstable for lower crime rates.

The $RECC$ shows almost no concentration under the varying scenarios with different crime rates. Although it is not a smooth function and presents some noise, it shows under varying crime rates that crime is not concentrated in the simulated populations.

Thus, the $RECC$ is a metric that gives consistent units of observation, which does not depend on the population size, on the threshold considered, on the scale of the spatial units and that does not simply ignore 80\% of the units.

\section{Conclusions and discussion}

\subsection{Are we just measuring randomness?}

Let's say, for instance, that we roll a fair 6-sided dice millions of times. Then, by the law of large numbers, we expect to observe, each side, roughly with the same frequency. And this is true and we should observe each side with the same frequency when rolling the dice millions of times. However, what happens if we roll the dice only 6 times? If we roll the dice only 6 times, then with a probability smaller than 2\% we will observe the six different sides of the dice and with a probability higher than 98\%, there will be at least one face that repeats and at least one side that never shows, thus, 98\% of the times we will observe some concentration, but this is simply the result of randomness combined with the low frequency of rolls.

In the case of crime, a similar situation occurs but, unfortunately, it is often ignored when measuring its concentration. With a large number of individuals or spatial units (street segments, for instance) and a small number of crimes, there will always be crime-free people or crime-free regions \cite{CrimeConcentrationRecommendations}, which as a result will show that crime is highly concentrated \cite{VancouverConcentration} but much of this apparent concentration would result also in the case of a random (uniform) distribution \cite{CrimeConcentrationVaryingCitySize}.

\subsection{Comparing metrics, comparing observations and comparing time periods}

Although far from having perfect metrics for the concentration of crime, three aspects (stability, comparability and independence of size and crime rates) are needed to accept a metric as the way in which the concentration of crime should be measured and compared. Here, a test against varying crime rates was constructed, but the same test could be used for detecting the impact of a varying population size.

\subsection{Comparing to randomness}

Much of the apparent concentration could be the result of a random (uniform) distribution \cite{CrimeConcentrationVaryingCitySize} and so a test against a uniform distribution of crime needs to be considered. In the case of individual victimisation, comparing the number of crimes suffered by each person under a uniform distribution of crime rates, is a simple test considering the number of crimes suffered by each person as a Poisson distribution with rate equal to the crime rate $c$ (see for instance \cite{PoissonRandomRepeats} and \cite{PoissonInCrime} for similar uses of the Poisson distribution in the context of crime science). By taking into account the number of crimes suffered by the population and comparing against random (Poisson) distributions with a constant rate $c$, then it is possible to compare the expected number of individuals who suffer zero crimes, one crime, two crimes and so on \cite{ConcentrationOfCrimeRECC}.

\bibliographystyle{unsrt}

\end{document}